\documentclass[dvips]{article}
\usepackage{icrctc07}

\title{Focal Plane Instrumentation of VERITAS}
\shorttitle{VERITAS camera}
\authors{T. Nagai$^{1}$, R. McKay$^{1}$, G. Sleege$^{1}$, D. Petry$^{2}$ for the VERITAS collaboration$^{3}$}
\shortauthors{T. Nagai for the VERITAS collaboration}
\afiliations{$^{1}$Department of Physics and Astronomy, Iowa State University Ames, IA 50011\\ $^{2}$Max-Planck-Institut f\"{u}r extraterrestrische Physik, Giessenbachstrasse, 85748 Garching, Germany\\  $^{3}$For full author list see G. Maier, {\it ``Status and Latest results of VERITAS''}, these proceedings.}
\email{nagai@iastate.edu}

\abstract{VERITAS is a new atmospheric Cherenkov imaging telescope
  array to detect very high energy gamma rays above 100 GeV. The array
  is located in southern Arizona, USA, at an altitude of 1268m above
  sea level. The array consists of four 12-m telescopes of
  Davies-Cotton design and structurally resembling the Whipple 10-m
  telescope. The four focal plane instruments are equipped with
  high-resolution (499 pixels) fast photo-multiplier-tube (PMT)
  cameras covering a 3.5 degree field of view with 0.15 degree pixel
  separation. Light concentrators reduce the dead-space between PMTs
  to 25$\%$ and shield the PMTs from ambient light. The PMTs are
  connected to high-speed preamplifiers allowing operation at modest
  anode current and giving good single photoelectron peaks in situ. 
  Electronics in the focus box provides real-time
  monitoring of the anode currents for each pixel and ambient
  environmental conditions. A charge injection
  subsystem installed in the focus box allows daytime testing of the
  trigger and data acquisition system by injecting pulses of variable
  amplitude and length directly into the photomultiplier
  preamplifiers. A brief description of the full VERITAS focal plane
  instrument is given in this paper.}

\begin{document}
\maketitle

\section{Introduction}

VERITAS (Very Energetic Radiation Imaging Telescope Array System) is a
successor of the pioneer imaging atmospheric Cherenkov telescope
(IACT), the Whipple 10-m telescope that was used to discover TeV
emission from a number of galactic and extra-galactic objects
including the Crab Nebula \cite{1989ApJ...342..379W}, the ``standard
candle'' in TeV astronomy. The array consists of four
telescopes giving stereoscopic views of showers, significantly
increasing gamma-ray flux sensitivity, angular resolution and energy
resolution relative to the stand-alone Whipple telescope. To fully
realize the capabilities of the four VERITAS telescopes, each is
equipped with a low-noise, high-resolution focal-plane instrument
coupled to a fast FADC-based data-acquisition system housed in a
nearby trailer.  The trigger system for the array has three tiers as
described elsewhere~\cite{jamie.fadc} \cite{2006APh....25..391H}.  

Each camera has 499 photomultiplier (PMT) pixels spanning a total
aperture of 3.5$^{\circ}$ with 0.15$^{\circ}$ pixel
spacing. Figure~\ref{camera.fig} shows the camera face with the light
concentrator plate (see below) installed. The pixels are supported
by a hexapod structure, allowing precise adjustment of the
position and tilt of the focal plane. The camera is designed to
operate at a trigger level of only 4 photoelectrons in the presence of
night-sky noise, the dominant background.  The relative electronic background
is reduced through the use of low-noise preamplifiers in the base of
the PMTs that boost the signal before it travels to the
data-acquisition electronics in the nearby trailers.  Because of this 
preamplification, the PMTs can be operated with reduced 
high voltage, that allows 
the array to be operated during partial moon night without damage to
the PMTs.  The remainder of this paper gives a brief description of
each of the instrument components and its performance.

\begin{figure}
\begin{center}
\includegraphics[width=0.38\textwidth]{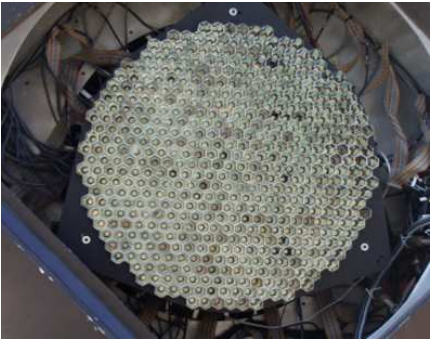}	
\end{center}
\caption{A VERITAS camera with 499 pixels. The light-collection cones 
in front of the PMTs reduce inter-tube dead space and shield against 
background light.}\label{camera.fig}
\end{figure}

\section{Photomultipliers and Preamplifiers}
\begin{figure*}
\begin{center}
\includegraphics[width=0.70\textwidth]{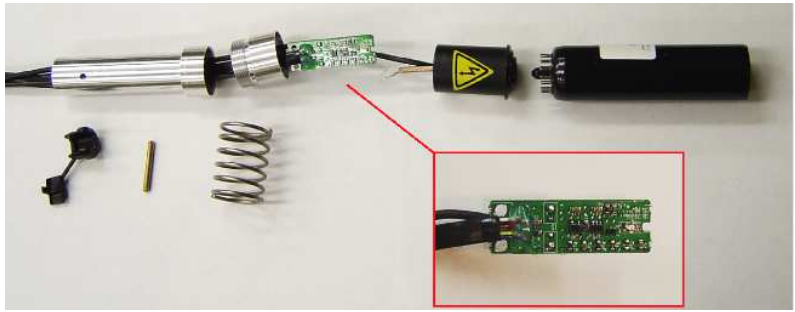}	
\end{center}
\caption{VERITAS pixel before assembly. The two-stage preamplifier is
  shielded in an aluminum shell. Two op-amps on the circuit board can be
  seen in the enlarged preamplifier image.}\label{pixel.fig}
\end{figure*}

Figure \ref{pixel.fig} shows the main components of a VERITAS
pixel. The photon sensors are Photonis XP2970/02 29mm diameter
PMTs with 10 gain stages, currently operated at a gain of $\sim 2 \times
10^{5}$. These have a typical quantum efficiency about 25$\%$
at wavelengths relevant for Cherenkov images (about 320 nm).

\begin{figure}[t]
\begin{center}
\includegraphics[width=0.38\textwidth]{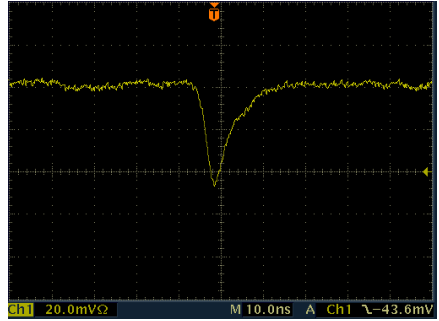} 
\end{center}	
\caption{Typical Cherenkov pulse with normal operating high voltage.}\label{pulse.fig}
\end{figure}

As illustrated in the blowup in the figure, a custom-built
preamplifier is installed in each PMT base. The main purpose
of the preamplifier is to boost the signal before its journey through
cables to the FADC's in the electronics trailer, thereby improving the
ratio of signal to electronic noise.  The preamplifier has a 
bandwidth of 300 MHz, in order to reproduce pulse shapes for fast 
  Cherenkov pulse rise time $\sim2.5$ ns (Figure \ref{pulse.fig}) and
good low-frequency response, preserving signal shapes down to 21 kHz.
Under normal operating conditions, the PMT gain combined with the
preamplifier amplification factor of 6.6 gives a single photoelectron
pulse height of 2.4 mV after 140 ft of coaxial cable (RG-59) and a
dynamic range of 0 to -2.2~V matched to the input of the FADC-based
data-acquisition system. The preamplifier also provides a direct DC
output for anode-current monitoring purposes as described in the next
section.

\begin{figure}
\begin{center}
\includegraphics[width=0.43\textwidth]{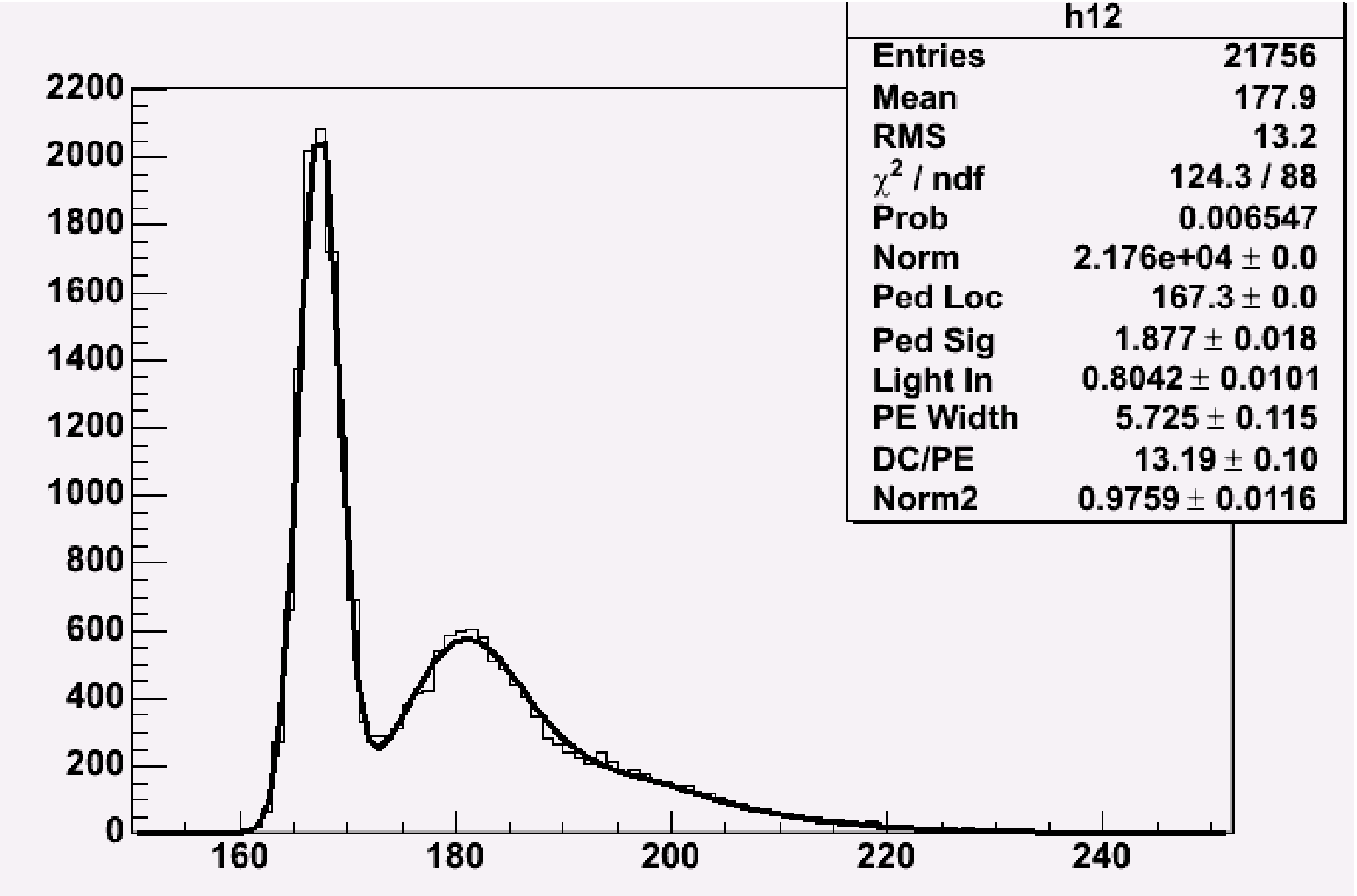} 
\end{center}	
\caption{The single photoelectron peak can be clearly seen at just over 180
digital counts with the increased gain by a factor of 3.}\label{singlepe.fig}
\end{figure}

With this quiet system we can recognize single photoelectron peaks
under standard operation conditions. However, to obtain well-defined 
single photoelectron peaks for more accurate calibration, we normally increase
the gain by a factor of 3 as shown in
Figure~\ref{singlepe.fig} where the single photoelectron peak is
clearly visible at just over 180 digital counts.

By modifying the mechanical and electrical connections between PMTs  
and bases, and by testing and selection,  we have obtained a percentage greater  
than 99$\%$ of fully functional installed pixels.


\section{Current Monitor Subsystem} 

It is important to monitor PMT anode currents in real time to protect
the tubes from transient light sources such as bright stars moving
through a tube's field of view or lights near the telescope accidentally
turned on. It is also important in detecting long-term changes in tube 
performance.  This is accomplished with custom-designed electronics inside the camera box
connected to electronics in the adjacent trailer through a fiber-optic
link.

Inside the camera box there are 16 FPGA-controlled circuit boards
daisy chained via a fast serial bus.\footnote{The design of these
  Field Programmable Gate Array (FPGA) boards is patented and the
  number of boards can be expanded up to 256. US Patent No. 6,717,514.}  Each board has 32
identical input channels multiplexed to one of the four 8-bit FADCs\footnote{Note that
  this FADC is {\it not} the FADC in the data-acquisition chain.}
operating at 50 MHz. The output is 1 digital count per 0.5 $\mu$A
giving a dynamic range of 0 to 127 $\mu$A.  Each of the channels can
be used for reading and transmitting either PMT anode currents or
environmental sensor outputs.

\begin{figure}[b]
\begin{center}
\includegraphics[width=0.38\textwidth]{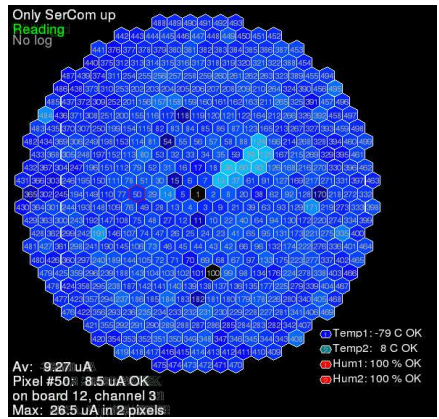} 
\end{center}	
\caption{VDCMON software screen shot. Anode currents from each 
PMT are shown as color-coded pixel map. The central pixel is removed for
telescope pointing procedure at the moment.}\label{vdcs}
\end{figure}

The channels are read, displayed and recorded by a computer in the
electronics trailer (Figure~\ref{vdcs}).  If the anode current of a PMT exceeds a preset
threshold, the PMT voltage is automatically reduced.  
 
\section{Charge Injection Subsystem} 

\begin{figure}[t]
\begin{center}
\includegraphics[width=0.38\textwidth]{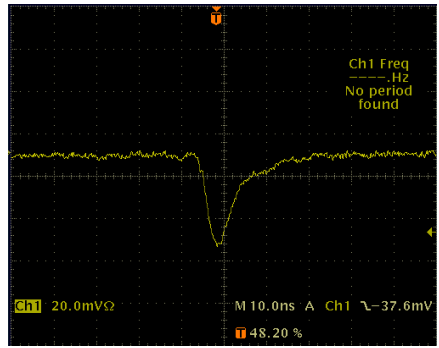} 
\end{center}	
\caption{Typical artificial pulse from the charge injection subsystem. 
The pulse resembles real Cherenkov pulses (see Figure \ref{pulse.fig}.)}\label{QI}
\end{figure}

This subsystem, installed inside the camera box, allows testing and
calibration the full data-acquisition chain following the PMTs.  This
is particularly convenient for working on the data-acquisition system
under non-observing conditions, e.g., during the day.  The
subsystem consists of a central charge injection circuit board inside
each camera that is fanned out to individual PMT channels.  The charge
injection circuit utilizes a programmable pulse generator (PPG) that
generates pulses (Figure \ref{QI}) with an adjustable frequency
ranging from 1 Hz to 1 MHz. The pulse height can be varied by 85dB,
and the output pulse width is also adjustable from 1 ns to 10 ms.
The generator is connected to fanout/mask boards attached to each of
the 16 current monitor boards.

The fanout/mask boards direct the pulses to particular pixels selected
by the operator.  The subsystem can then be used for detailed
studies of triggering efficiency using different triggering patterns,
crosstalk, diagnostics to find mis-wired/mislabeled channels and the
ability to inject pseudo-Cherenkov images into the electronics
system.

\section{Light Concentrators} 

\begin{figure}
\begin{center}
\includegraphics[width=0.38\textwidth]{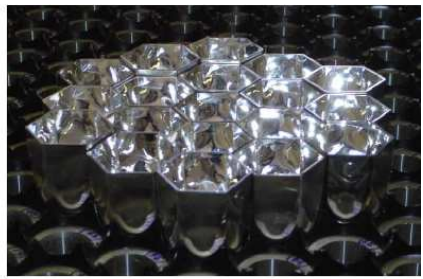}
\end{center}	
\caption{A group of light cones laid on a Delrin light-cone plate. 
Each cone is made of molded half-cones.}\label{cones.fig}
\end{figure}

A light concentrator plate is installed on each camera to reduce dead
space between PMTs increasing signal light collection and to limit
the acceptance angle of the pixels to the solid angle subtended by
the telescope thereby reducing background light.  It consists of 499
molded plastic cones glued onto a machined Delrin plate.  (Figure
\ref{cones.fig} shows a small group of light cones sitting on the
plate.) The inner surface of each cone has an evaporated aluminum
coating with a protective overlayer \footnote{The coating is done by
  Evaporated Coatings, Inc. (ECI) with coating $\#$801p. \\
(http://www.evaporatedcoatings.com/)} giving greater than $85
\%$ reflectivity above 260 nm.  The cone shape is a hybrid design with
a hexagonal entrance window evolving to a Winston cone at the exit.
The effect of the cones is to increase the geometrical light
collection efficiency at the photocathodes from 55$\%$ to 75$\%$ \cite{lightcone}.

\section{Acknowledgements}
VERITAS is supported by grants from the U.S. Department of Energy, the U.S. 
National Science Foundation and the Smithsonian Institution, by NSERC in
Canada, by PPARC in the U.K. and by Science Foundation Ireland. The authors 
also acknowledge support from Iowa State University.


\bibliography{icrc0757}
\bibliographystyle{nature}

\end{document}